# An Axis Symmetric 2D Description of the Dynamics of a Single Crystal Micro Fiber Growth from the Melt by Micro- Pulling- Down Method. Part.1

Agneta M. BALINT
Department of Physics
West University of Timisoara
Timisoara, Romania
e-mail: agneta.balint@e-uvt.ro

Stefan BALINT
Department of Computer Science
West University of Timisoara
Timisoara, Romania
e-mail: stefan.balint@e-uvt.ro

*Abstract*—**In this paper the first part of an axis symmetric 2D description of a single crystal micro fiber growth process by micro-pulling–down (μ-PD) method is presented. The description concerns the following aspects: the free surface equation and the pressure difference across the free surface (section 2); limits of the pressure difference p across the free surface (section 3); the fiber radius change rate, due to the change of the pressure difference p across the free surface , and the fiber radius size control (section 4);the static stability of the free surface (section 5). The above aspects are illustrated numerically in the case of the growth of a Si micro-fiber of radius 0.0001[m] by using the Maple 17 software. This description can be helpful in the better understanding of the growth process and in the automation of manufacturing.**

## I. INTRODUCTION

The near one-dimensional single crystal fibers have attracted some attention on the applications of optical and electronic devices [1-3]. There are several methods for growing fiber crystals [4]. The micro-pulling down (μ-PD) process, a variant of the inverse edge-defined film-fed growth, developed by Fukuda's laboratory in Japan [5-9], has been shown promising in producing single crystal fibers with good diameter control and concentration uniformity. According to [10], for this process some simple theoretical analysis for the operation limits [11] and solute distribution [12, 13] have been performed and no detailed modeling has been conducted. In [14] the meniscus shape and size, appearing in μ-PD were evaluated in function of pressure difference, which incorporates the hydrostatic pressure of the melt column situated behind the meniscus and the pressure in front of the free surface of the meniscus. In [15] some results from [14] were used in the analysis of the effect of the pressure difference across the free surface on the shape and size of the meniscus in an NdYAG microfiber growth from the melt by μ-PD method.

## II. THE FREE SURFACE EQUATION

The free surface of the meniscus (see Fig.1) is described mathematically by the Young-Laplace equation [16,17]:

$$\gamma \cdot \left( \frac{1}{R_1} + \frac{1}{R_2} \right) = P_a - P_b \quad (1)$$

Here: $\gamma$ is the melt surface tension; $1/R_1$, $1/R_2$ denote the main normal curvatures of the free surface at a point M of the free surface; $P_a$ is the pressure in front of the free surface; $P_b$ is the pressure behind the free surface. (Fig.1). The pressure $P_b$ behind the free surface is the sum of the hydrodynamic pressure $p_m$ in the meniscus melt (due to the thermal convection), the Marangoni pressure $p_M$ (due to the Marangoni convection) and the hydrostatic pressure of the melt column behind the free surface equal to $\rho \cdot g \cdot (z + h_m)$ (see Fig.1). Here: $\rho$ denotes the melt density; $g$ is the gravity acceleration; $z$ is the coordinate of M with respect to the $Oz$ axis, directed vertically downwards; $h_m$ denotes the melt column height between the horizontal crucible melt level and the shaper bottom level (Fig.1). The pressure difference $P_a - P_b$ across the free surface, denoted usually by $\Delta p$, according to the above considerations is:

$$\Delta p = P_a - p_m - p_M - \rho \cdot g \cdot (z + h_m). \quad (2)$$

The part $p$ of $\Delta p$ is given by:

$$p = -P_a + \rho \cdot g \cdot h_m + p_m + p_M \quad (3)$$

is independent of the spatial coordinate $z$ (it can depend on the moment of time $t$ ) and the major part of $p$ is $-P_a + \rho \cdot g \cdot h_m$ because the sum $p_m + p_M$ in general is small with respect to $-p_g + \rho \cdot g \cdot h_m$. For this reason thereafter is assumed that $p$ is given by:

$$p = -P_a + \rho \cdot g \cdot h_m \quad (4)$$

Remark that $p$ given by (4) can be controlled by $P_a$ and $h_m$.

With this approximation the Young-Laplace free surface equation (1) can be written as:

$$\gamma \cdot \left( \frac{1}{R_1} + \frac{1}{R_2} \right) = -\rho \cdot g \cdot z - p \quad (5)$$

Therefore, for an axis-symmetric meniscus free surface the differential equation of the meridian curve is given by:

$$z'' = -\frac{\rho \cdot g \cdot z + p}{\gamma} \left[ 1 + (z')^2 \right]^{3/2} - \frac{1}{r} \cdot \left[ 1 + (z')^2 \right] \cdot z'$$

$$R_c \leq r \leq R_d \quad (6)$$

where $R_c$ is the crystal radius, $R_d$ is the shaper radius and $R_h$ is the capillary channel radius.

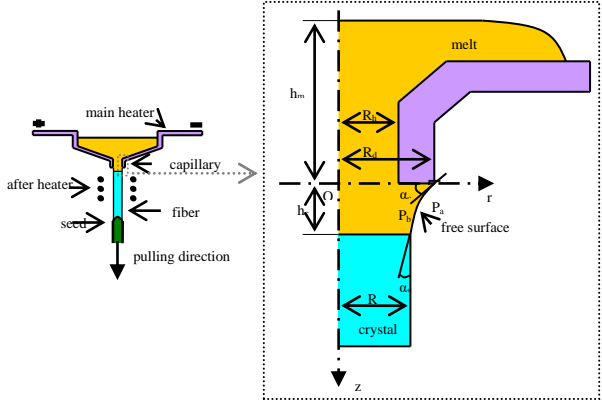

Figure 1. Schematic diagram of the µ-pulling-down fiber

The function $z(r)$, describing the meridian curve profile, verifies the following boundary conditions:

$$z'(R_c) = -\tan(\pi/2 - \alpha_g); \quad z(R_c) = h_c > 0 \quad (7)$$

$$z'(R_d) = -\tan \alpha_c; \quad z(R_d) = 0 \quad (8)$$

$$z(r) \text{ is strictly decreasing on } [R_c, R_d]. \quad (9)$$

The first condition in (7) expresses that at the three phase point $(R_c, h_c)$, where the thermal conditions for solidification have to be realized ($h_c$ is the crystallization front level), the angle, between the tangent line to the meridian curve of the free surface and the vertical, is equal to the growth angle $\alpha_g$. If this condition is satisfied during the whole process, then the fiber diameter is constant.

The second condition in (7) expresses that the coordinate of the crystallization front level is $h_c > 0$.

The first condition in (8) expresses that at the point $(R_d, 0)$, where the meridian curve touches the outer edge of the shaper, the catching angle is equal to $\alpha_c$.

The second condition in (8) expresses that at the point $(R_d, 0)$ the meridian curve is fixed to the outer edge of the shaper.

Condition (9) expresses that the meniscus shape is relatively simple.

Beside the conditions (6), (7), (8) and (9) the function $z(r)$ describing the meridian curve profile, has to minimize the free energy functional of the melt column behind the free surface given by:

$$I(z) = \int_{R_c}^{R_d} \left\{ \gamma \cdot [1 + (z')^2]^{1/2} - \frac{1}{2} \cdot \rho \cdot g \cdot z^2 - p \cdot z \right\} \cdot r \cdot dr \quad (10)$$

This last condition is called the static stability of the axis symmetric free surface and in real world only statically stable capillary free surfaces exists..

### III. LIMITS OF THE PRESSURE DIFFERENCE $p$

If $p$ is arbitrary, then it can happen that there is no function $z(r)$ defined for $r \in [R_c, R_d]$ which verifies (6) - (9) and supplementary $z''(r) > 0$. This means that there is no convex static axis-symmetric meniscus. The following statement is a necessary condition for the existence of a function $z(r)$ having the properties (6) - (9) and $z''(r) > 0$ for $r \in [R_c, R_d]$.

**Statement 1** *For $\alpha_c + \alpha_g < \pi/2$, in order to have an axis-symmetric meniscus free surface with convex meridian curve on $[R_c = R_d/n, R_d]$ the pressure across the free surface has to be in the range given by:*

$$\gamma \cdot \frac{\alpha_c + \alpha_g - \pi/2}{R_d} \cdot \frac{n}{n-1} \cdot \cos \alpha_c - \rho \cdot g \cdot R_d \cdot \frac{(n-1)}{n} \cdot \tan(\pi/2 - \alpha_g)$$

$$-\frac{\gamma}{R_d} \cdot n \cdot \cos \alpha_g \leq p \leq \gamma \cdot \frac{\alpha_c + \alpha_g - \pi/2}{R_d} \cdot \frac{n}{n-1} \cdot \sin \alpha_g - \frac{\gamma}{R_d} \cdot \sin \alpha_c$$

$$(11)$$

*where $n = R_d / R_c > 1$.*

The proof of this statement can be found in the Appendix. In case of a Si microfiber for the following numerical data:
$R_d = 2 \times 10^{-4} [m]$; $\alpha_c = 0.523 [rad]$; $\alpha_g = 0.192 [rad]$;
$\rho = 2.58 \times 10^3 [kg/m^3]$; $g = 9.81 [m/s^2]$; $\gamma = 0.720 [N/m]$
and $R_c = 1 \times 10^{-4} [m]$ (the value of the contact angle $\alpha_c = 0.523 [rad]$ is the smallest admissible value for silicon according to Yang et al. [24]) the pressure limits appearing in (11) (Statement 1) are denoted by:

$$L_1(n) = \gamma \cdot \frac{\alpha_c + \alpha_g - \pi/2}{R_d} \cdot \frac{n}{n-1} \cdot \cos \alpha_c -$$

$$\rho \cdot g \cdot R_d \cdot \frac{(n-1)}{n} \cdot \tan(\pi/2 - \alpha_g) - \frac{\gamma}{R_d} \cdot n \cdot \cos \alpha_g \quad (12)$$

$$L_2(n) = \gamma \cdot \frac{\alpha_c + \alpha_g - \pi/2}{R_d} \cdot \frac{n}{n-1} \cdot \sin \alpha_g - \frac{\gamma}{R_d} \cdot \sin \alpha_c \quad (13)$$

For $R_c = R_d/2$ the value of $n$ is $n = R_d/R_c = 2$ and the pressure limits are $L_1(2) = -12418.77818\ [Pa]$ ; $L_2(2) = -2973.93043\ [Pa]$. So the values of $p$ have to be in the range $[-12418.77818\ ;\ -2973.93043][Pa]$. In practice the value of $p = -P_a + \rho \cdot g \cdot h_m$ can be controlled through the values of $P_a$ and $h_m$, but the range where the value of $p$ has to be is large and we don't know which value of $p$ is appropriate. Moreover, is not sure that there exists a value of $p$ in the above range for which an axis symmetric meniscus with convex meridian curve exists (the condition (11) is only necessary).

In order to answer these questions the following initial value problem:

$$\begin{cases} \dfrac{dz}{dr} = -\tan\alpha \\ \dfrac{d\alpha}{dr} = \dfrac{\rho \cdot g \cdot z + p}{\gamma} \cdot \dfrac{1}{\cos\alpha} + \dfrac{1}{r} \cdot \tan\alpha \\ z(R_d) = 0,\quad \alpha(R_d) = \alpha_c \end{cases} \quad (14)$$

has to be solved numerically for different values of $p$ in the obtained range. In this way it is found that the value of $p$ for which $R_c = 1 \times 10^{-4}[m]$ and $\alpha(R_c) = 1.37879632\ [rad]$ is $p = -8079\ [Pa]$. For this value of $p$ the meniscus height is $h_c = 1.907095 \times 10^{-4}[m]$. The meridian curve of the meniscus is presented in Fig.2.a and $\alpha(r)$ in Fig.2.b.

At this point it has to be noted that the downward orientation of the OZ axis in Fig.1. and the upward orientation of the OZ axis in figure Fig.3a are opposite. For that this last figure has to be rotated with 180 degrees around the OX axis in order to obtain the meridian curve shape as it is presented in Fig.1.

Computation reveals that in the considered range there is no other value of $p$ such that for $R_c = 1 \times 10^{-4}[m]$

$\alpha(R_c) = \pi/2 - \alpha_g = 1.37879632\ [rad]$

**Remark.** The same computation reveal that for different values of $p$ in the range $[-12418, -8059][Pa]$ there exist different values of $R_c(p)$ for which $\alpha(R_c(p)) = \pi/2 - \alpha_g = 1.37871\ [rad]$
More than that, when $p$ increases from the value $-12418\ [Pa]$ to the value $-8059\ [Pa]$, then $R_c(p)$, having the above property, decreases from the value $R_c(-12418) = 1.5976 \times 10^{-4}[m]$ to the value $R_c(-8059) = 9 \times 10^{-5}[m]$ and $z(R_c(p)) = h_c(p)$ increases from the value $h_c(-12418) = 5.920033 \times 10^{-5}[m]$ to the value $h_c(-8059) = 2.390507 \times 10^{-4}[m]$. This means that to any $p$ in the range $[-12418, -8059][Pa]$ correspond a unique axis symmetric 2D meniscus with convex meridian curve $z = z(r, p)$, but for only one of them, namely for $p = -8079\ [Pa]$ the following equalities hold $R_c(p) = 1 \times 10^{-4}[m]$  $\alpha(R_c(p)) = 1.37871\ [rad]$.

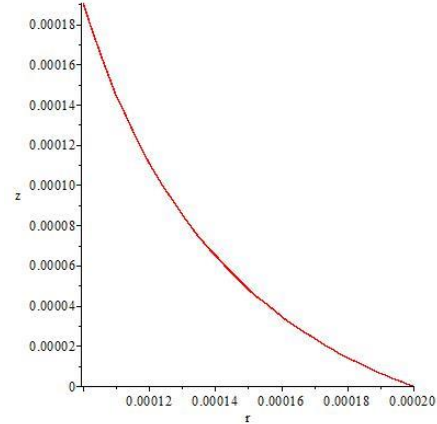

Figure 2a. Meniscus meridian curve shape $z = z(r)$
$p = -8079\ [Pa]$, $R_c = 1 \times 10^{-4}[m]$, $h_c = 1.907095 \times 10^{-4}[m]$.

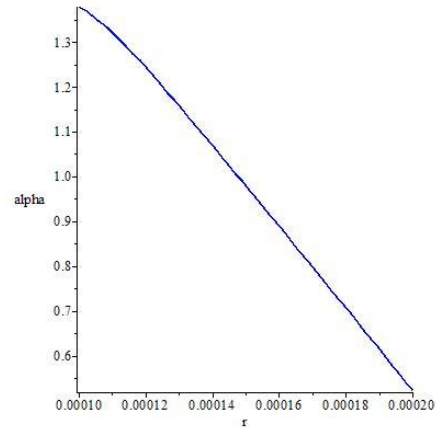

Figure 2.b Variation of $\alpha(r)$.
$\alpha(R_c, p) = 1.378710\ [rad]$

Computation reveals also that for any $p$ in the range $[-8100, -8059][Pa]$ it makes sense to compute $\alpha(0.0001, p)$ and $z(0.0001, p)$.
For $p = -8100\ [Pa]$ , $\alpha(0.0001, p) = 1.3888455729\ [rad]$ and $z(0.0001, p) = 0.000194301384\ [m]$ while for $p = -8059\ [Pa]$ , $\alpha(0.0001, p) = 1.36936083\ [rad]$ and $z(0.0001, p) = 0.000187574\ [m]$

The angle $\alpha(0.0001, p)$ considered above is the angle made by the line tangent to the meniscus (corresponding to $p$) at the point $(0.0001, z(0.0001, p))$ with the horizontal axis. This angle is equal to $\pi/2 - \alpha_g = 1.37871023884957\ [rad]$ if and only if $p = -8079\ [Pa]$. For $p$ in the range

$[-8100,-8079)\,[Pa]$ the angle $\alpha(0.0001,p)$ satisfies $\alpha(0.0001,p) > \pi/2 - \alpha_g = 1.37871023884957\,[rad]$ while for $p$ in the range $(-8079,-8059]\,[Pa]$ it satisfies $\alpha(0.0001,p) < \pi/2 - \alpha_g = 1.37871023884957\,[rad]$.

For $p$ different from $-8709\,[Pa]$ the difference
$$\alpha(0.0001,p) - (\pi/2 - \alpha_g) =$$
$$\alpha(0.0001,p) - 1.37871023884957\,[rad]$$
represents the deviation from the growth angle, due to the deviation of the pressure difference $p$ from the value $p = -8079\,[Pa]$. For $p$ in the range $[-8100,-8079)\,[Pa]$ the deviation is strictly positive and for $p$ in the range $(-8079,-8059]\,[Pa]$ the deviation is strictly negative.

## IV. EQUATION OF THE FIBER RADIUS CHANGE RATE DUE TO THE CHANGE OF THE PRESSURE DIFFERENCE AND FIBER RADIUS SIZE CONTROL

Starting from the condition of growth angle constancy, according to [17], the following equation of the fiber radius change rate, due to the pressure difference perturbation $p$, is obtained:
$$dR_c/dt = -v\tan(\alpha(0.0001,p) - (\pi/2 - \alpha_g)) \quad (15)$$
$$R_c(0) = 0.0001$$

Here $R_c(t)$ is the fiber radius at the moment of time $t$, $v$ (the pulling rate) for the moment is a strictly positive constant and $\alpha(0.0001,p)$ is the angle made by the line tangent to the perturbed meniscus at the point of coordinates $(0.0001, z(0.0001,p))$, with the horizontal axis.

If $p = -8079\,[Pa]$, then the initial value problem (15) becomes:
$$dR_c/dt = 0 \quad (16)$$
$$R_c(0) = 0.0001$$

For $p = -8100\,[Pa]$ we have $\alpha(0.0001,p) = 1.3888455\,[rad]$ and the initial value problem (15) becomes:
$$dR_c/dt = -v\tan(\alpha(0.0100492)) \quad (17)$$
$$R_c(0) = 0.0001$$

For $p = -8059\,[Pa]$ we have $\alpha(0.0001,p) = 1.3693608\,[rad]$.
So
$$-v\cdot\tan(\alpha(0.0001,p) - (\pi/2 - \alpha_g)) = -v\cdot\tan(-0.0094355)$$
and the initial value problem (15) becomes:
$$dR_c/dt = v\tan(0.0094355) \quad (18)$$
$$R_c(0) = 0.0001$$

For $v = 0.0001\,[m/s]$ the solutions of the initial value problems (16), (17) and (18) are represented on Fig.3 with green, blue and red respectively.

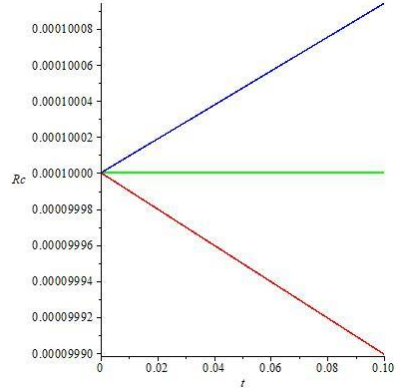

Figure 3. The fiber radius change for $v = 0.0001\,[m/s]$ and $p_1 = -8079\,[Pa]$ (green), $p_2 = -8100\,[Pa]$ (red), $p_3 = -8059\,[Pa]$ blue, respectively.

On Fig.3 can be seen that:
- if in the first $0.1\,[s]$ $p$ is constant equal to $-8079\,[Pa]$, then the fiber radius is constant equal to $R_c = 0.0001\,[m]$ (color green)
- if at $t = 0$, $p$ decreases instantaneously from $-8079\,[Pa]$ to $-8100\,[Pa]$, then the fiber radius decreases in $0.1\,[s]$ from $R_c = 0.0001\,[m]$ to $R_c{'} = 0.00009989950\,4171\,[m]$ (color red)
- if at $t = 0$, $p$ increases instantaneously from $-8079\,[Pa]$ to $-8059\,[Pa]$, then the fiber radius increases in $0.1\,[s]$ from $R_c = 0.0001\,[m]$ to $R_c{''} = 0.0001000943\,578002\,[m]$ (color blue).

In the two cases when the radius $R_c = 0.0001\,[m]$ varies due to the pressure perturbation the problem which appear is: „what is the procedure which has to be applied in order to recover the radius $R_c = 0.0001\,[m]$"?

In the following two procedures are presented:
**a).** the recovery of the value $R_c = 0.0001\,[m]$ starting from the value $R_c{'} = 0.00009989950\,4171\,[m]$.
**b).** the recovery of the value $R_c = 0.0001\,[m]$ starting from the value $R_c{''} = 0.0001000943\,578002\,[m]$.

Both procedures are based on the pressure difference value modification at the moment of time $t = 0.1\,[s]$.

**a).** The first step in this procedure is the determination of the static meniscus for which the radius is $R_c{'} = 0.00009989950\,4171\,[m]$. This means in fact the determination of the pressure difference $p'$ for which the stable meniscus radius is $R_c{'}$.

Computation shows that we have to take $p' = -8077\,[Pa]$ and the computed results for this $p'$ are :

$R_c' = 0.0000998995\ 04171\,[m]$

$z(R_c', p') = h_c' = 0.0001909474\ 878793\,[m]$

$\alpha(R_c', p') = 1.3787988372\ 1489\,[rad]$.

The second step in this procedure is to find a perturbation $p$ of $p'$ for which the radius $R_c'$ increases. For that we have to look for a value of $p$ which is more than $p' = -8077\,[Pa]$. For example if we take $p = -8059\,[Pa]$ computation reveals that : for $R_c' = 0.0000998995\ 04171\,[m]$ we have $z(R_c', p) = 0.0001\,8806729150\ 7972\,[m]$ and $\alpha(R_c', p') = 1.3697254484\ 3397\,[rad]$. Therefore, for $p = -8079\,[Pa]$ starting from $R_c' = 0.0000998995\ 04171\,[m]$ at the moment of time $t = 0{,}1\,[s]$ the increase of $R_c'$ is described by the solution of the initial value problem

$$dR_c'/dt = -0.0001 \cdot \tan(\,-0.009070879\,) \quad (19)$$
$$R_c'(0.1) = 0.00009989950417$$

The increase and the complete process of decrease and recovery of $R_c = 0.0001\,[m]$ is presented in Fig.4.a. and Fig.4.b.

**b).** The first step in this procedure is the determination of the static meniscus for which the radius is $R_c'' = 0.0001000943\ 578002\,[m]$. This means in fact the determination of the pressure difference $p''$ for which the stable meniscus radius is $R_c''$. Computation shows that we have to take $p'' = -8080.15\,[Pa]$ and the computed result for this $p''$ are: $R_c'' = 0.0001000943\ 578002\,[m]$, $\alpha(R_c'', p'') = 1.3787988372\ 1489\,[rad]$, $z(R_c'', p'') = h_c'' = 0.0001904119\ 03269816\,[m]$.

The second step in this procedure is to find a perturbation $p$ of $p''$ for which the radius $R_c'' = 0.0001000943\ 578002\,[m]$ decreases. For that we have to look for a value of p which is less than p''. For example if we take $p = -8100\,[Pa]$ computation reveals that: for $R_c'' = 0.0001000943\ 578002\,[m]$ we have $z(R_c'', p) = 0.0001145146\ 33617455\,[m]$ and $\alpha(R_c'', p) = 1.3884539105\ 4896\,[rad]$. Therefore for the value $p = -8100\,[Pa]$ starting from $R_c'' = 0.0001000943\ 578002\,[m]$ at the moment of time $t = 0{,}1\,[s]$ the decrease of $R_c''$ is described by the solution of the initial value problem

$$dR''_c/dt = -0.0001 \cdot \tan(0.009657584\,) \quad (20)$$
$$R_c''(0.1) = 0.0001000943578002$$

The decrease and the complete process of increase and recovery of $R_c = 0.0001\,[m]$ is presented on Fig.5.

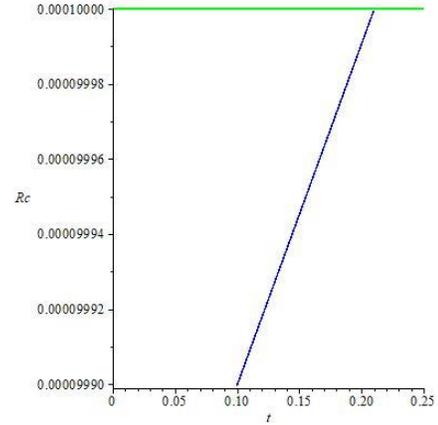

Figure 4.a. Increase of $R_c'(t)$ for $t$ in the range $[0.1, 0.21]\,[s]$.

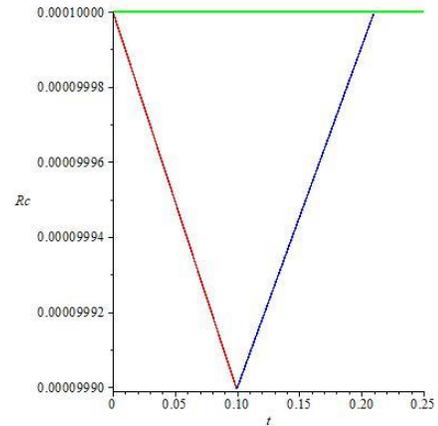

Figure 4.b. Decrease from $R_c = 0.0001\,[m]$ to $R_c' = 0.0000998995\ 04171\,[m]$ and recovery to $R_c = 0.0001\,[m]$

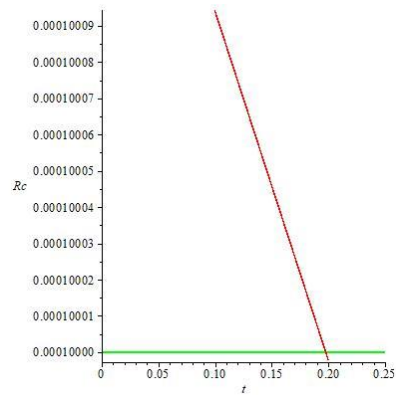

Figure 5 a. Decrease of $R_c''(t)$ for t in the range $[0.1, 0.2]\,[s]$

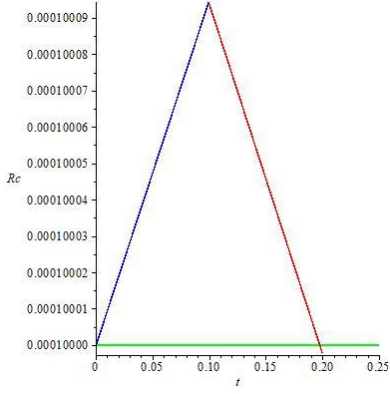

Figure 5 b. Increase from $R_c = 0.0001 \, [m]$ to $R_c'' = 0.0001000943\,578002 \, [m]$ and recovery to $R_c = 0.0001 \, [m]$

If $p = p(t)$ oscillates around the value $-8079 \, [Pa]$, then $\alpha(0.0001, p)$ oscillates around the value $\pi/2 - \alpha_g$. In this case the plate half thickness evolution is described by the solution of the initial value problem:

$$dR_c/dt = -v \tan(\alpha(0.0001, p(t)) - (\pi/2 - \alpha_g)) \quad (21)$$
$$R_c(0) = 0.0001$$

For $v = 0.0001 \, [m/s]$ and $\alpha(0.0001, p(t)) = \pi/2 - \alpha_g + \sin(600t)$ the fiber radius evolution during the first second can be seen in Fig.6a. On Fig.6b the plate half thickness evolution during the first second is presented for the already considered instantaneous perturbations and for the above considered oscillatory perturbation.

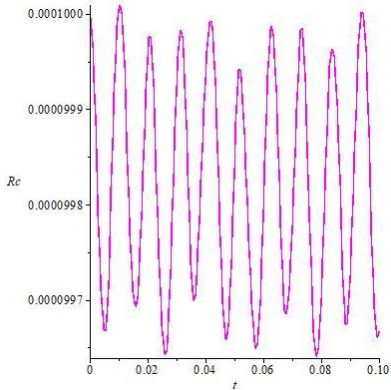

Figure 6.a. The fiber radius change for $\alpha(0.0001, p(t)) = \pi/2 - \alpha_g + \sin(600t)$

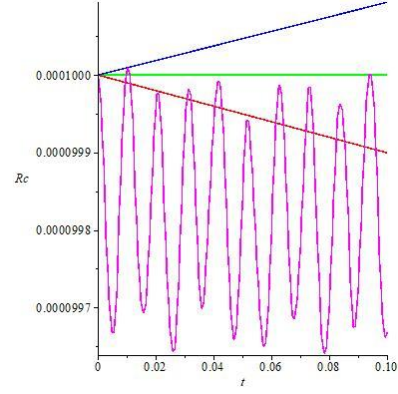

Figure 6.b. The fiber radius change for $v = 0.0001 \, [m/s]$ $p = -8079 \, [Pa]$ (green), $p = -8100 \, [Pa]$ (red), $p = -8059 \, [Pa]$ (blue) $\alpha(0.0001, p(t)) = \pi/2 - \alpha_g + \sin(600t)$ (magenta).

## V. STATIC STABILITY OF THE FREE SURFACE

This section deals with the static stability of the free surface [14]. The following statement is a sufficient condition of static stability of the free surface.

**Statement 2** If $n = R_d/R_c > 1$ satisfies

$$(n-1)/n^{1/2} < \pi \cdot \gamma^{1/2} \cdot \sin^{3/2} \alpha_g / R_d \cdot \rho^{1/2} \cdot g^{1/2} \quad (22)$$

then the axis-symmetric meniscus is stable, i.e. $z(r)$ minimizes the free energy functional $I(z)$.

If $n = R_d/R_c > 1$ satisfies

$$(n-1)/n^{3/2} > 2\pi \cdot \gamma^{1/2} \cdot \cos^{3/2} \alpha_c / R_d \cdot \rho^{1/2} \cdot g^{1/2} \quad (23)$$

then the axis-symmetric meniscus is unstable, i.e. $z(r)$ is not a minimum for the free energy functional $I(z)$.

In the case of the Si fiber considered in section 3 we have: $(n-1)/n^{1/2} = 0.7071067814$ and $\pi \cdot \gamma^{1/2} \cdot \sin^{3/2} \alpha_g / R_d \cdot \rho^{1/2} \cdot g^{1/2} = 7.198643944$. Therefore, the static free surface is stable.

## VI. CONCLUSIONS

In the framework of the presented description, the following quantities, related to the micro fiber growth process from the melt by pulling down method, can be computed:
- pressure difference across the free surface for create a stable static meniscus appropriate for the growth of a fiber having a prior given radius.
- pressure difference across the free surface for controlling the fiber radius size during the growth

# ANNEXE

## Proof of the Statement 1.

Let $\alpha_c + \alpha_g < \pi/2$, $R_c = R_d/n$, $1 < n$ and $z(r)$ defined for $r \in [R_c = R_d/n, R_d]$ which verifies (7), (12) - (14) and $z''(r) > 0$. The function defined as:

$$\alpha(r) = -\arctan(z'(r)) \quad \text{for } r \in [R_c = R_d/n, R_d]$$

verifies

$$\alpha'(r) = \frac{\rho \cdot g \cdot z(r) + p}{\gamma} \cdot \frac{1}{\cos(\alpha(r))} + \frac{1}{r} \cdot \tan(\alpha(r))$$

and the boundary conditions: $\alpha(R_d) = \alpha_c$, $\alpha(R_d) = \pi/2 - \alpha_g$.

Hence, by the mean value theorem, there exists $r' \in [R_c, R_d]$ such that the following equality holds:

$$p = \gamma \cdot \frac{\alpha_c + \alpha_g - \pi/2}{R_d - R_c} \cdot \cos\alpha(r') - \rho \cdot g \cdot z(r') - \frac{\gamma}{r'} \cdot \sin(\alpha(r'))$$

On the other hand, inequality $z''(r) > 0$ implies that the function $z'(r)$ is strictly increasing and by consequence the function $\alpha(r)$ is strictly decreasing. Therefore, $\alpha_c < \alpha(r') < \pi/2 - \alpha_g$. Now, by taking into account that $z(r') < 0$ and $\gamma \cdot \frac{\alpha_c + \alpha_g - \pi/2}{R_d - R_c} < 0$ the following relations hold:

$$p = \gamma \cdot \frac{\alpha_c + \alpha_g - \pi/2}{R_d - R_c} \cdot \cos\alpha(r') - \rho \cdot g \cdot z(r') - \frac{\gamma}{r'} \cdot \sin(\alpha(r')) <$$

$$\gamma \cdot \frac{\alpha_c + \alpha_g - \pi/2}{R_d - R_c} \cdot \cos(\frac{\pi}{2} - \alpha_g) - \gamma \cdot \frac{1}{R_d} \cdot \sin\alpha_c <$$

$$\gamma \cdot \frac{\alpha_c + \alpha_g - \pi/2}{R_d} \cdot \frac{n}{n-1} \sin(\alpha_g) - \frac{\gamma}{R_d} \cdot \sin(\alpha_c)$$

with $n = R_d/R_c > 1$.

So, the right hand side of the inequality (12) is proven.

In order to obtain the left hand side of (11) remark first the inequality:

$$p = \gamma \cdot \frac{\alpha_c + \alpha_g - \pi/2}{R_d - R_c} \cdot \cos\alpha(r') - \rho \cdot g \cdot z(r') -$$

$$\frac{\gamma}{r'} \cdot \sin(\alpha(r')) > \gamma \cdot \frac{\alpha_c + \alpha_g - \pi/2}{R_d - R_c} \cdot \cos\alpha_c -$$

$$\rho \cdot g \cdot z(R_c) - \frac{\gamma}{R_c} \cdot \cos(\alpha_g).$$

The term $-\rho \cdot g \cdot z(R_c)$ can be evaluated applying the mean value theorem for the function $z(r)$ on $[R_c, R_d]$. It follows that there exists $r'' \in [R_c, R_d]$ such that

$$z(R_d) - z(R_c) = (R_d - R_c)z'(r'') > (R_d - R_c)z'(R_d) =$$

$$-(R_d - R_c) \cdot \tan(\frac{\pi}{2} - \alpha_g)$$

$z(R_d) = 0$.

So, for $p$ the following relations hold:

$$p > \gamma \cdot \frac{\alpha_c + \alpha_g - \pi/2}{R_d - R_c} \cdot \cos\alpha_c - \rho \cdot g \cdot (R_d - R_c) \cdot \tan(\alpha_c) -$$

$$\frac{\gamma}{R_c} \cdot \cos(\alpha_g) > \gamma \cdot \frac{\alpha_c + \alpha_g - \pi/2}{R_d} \cdot \frac{n}{n-1} \cdot \cos\alpha_c -$$

$$\rho \cdot g \cdot R_d \cdot \frac{n-1}{n} \cdot \tan(\frac{\pi}{2} - \alpha_g) - \frac{\gamma}{R_d} \cdot n \cdot \cos(\alpha_g)$$

This means that the left hand side of (12) is proven.

## Proof of the Statement 2.

Since (7) is the Euler equation [27] Chapter 2 for the free energy functional (15), in this case it is sufficient to investigate the Legendre and Jacobi conditions [27] Chapter 8.

Consider for that the function

$$F(z, z', r) = \left\{ \gamma \cdot [1 + (z')^2]^{1/2} - \frac{1}{2} \cdot \rho \cdot g \cdot z^2 - p \cdot z \right\} \cdot r$$

and remark that the Legendre condition $\partial^2 F/\partial z'^2 > 0$ reduces to the inequality:

$$r \cdot \gamma \left[1 + (z')^2\right]^{-3/2} > 0$$

which is satisfied.

The Jacobi equation

$$\left[\frac{\partial^2 F}{\partial z^2} - \frac{d}{dr}\left(\frac{\partial^2 F}{\partial z \partial z'}\right)\right] \cdot \eta - \frac{d}{dr}\left[\frac{\partial^2 F}{\partial z'^2} \cdot \eta'\right] = 0$$

in this case becomes

$$\frac{d}{dr}\left(\frac{r \cdot \gamma}{[1 + (z_e')^2]^{3/2}} \cdot \eta'\right) + \rho \cdot g \cdot r \cdot \eta = 0$$

i). For obtaining the stability result remark that using (7), (12)-(14) and $z''(r) > 0$ for the coefficients of the Jacobi equation the following inequalities can be obtained:

$$r \cdot \gamma/[1 + (z')^2]^{3/2} \geq (R_d \cdot \gamma \cdot \sin^3\alpha_g)/n$$

$$\rho \cdot g \cdot r \leq \rho \cdot g \cdot R_d$$

Therefore the equation

$$\frac{d}{dr}\left(\frac{R_d}{n} \cdot \gamma \cdot \sin^3\alpha_g \cdot \varsigma'\right) + \rho \cdot g \cdot R_d \cdot \varsigma = 0$$

is a „Sturm type upper bound" ([28] Chapter 11) for the Jacobi equation. An arbitrary solution of the above „Sturm type upper bound equation" is given by

$$\varsigma(r) = A \cdot \sin(\omega \cdot r + \varphi)$$

Here $A$ and $\varphi$ are arbitrary real constants and $\omega^2 = \rho \cdot g \cdot n / \gamma \cdot \sin^3 \alpha_g$.

The half period of any non zero solution $\varsigma(r)$ is $\pi/\omega = \pi \cdot \gamma^{1/2} \cdot \sin^{3/2} \alpha_g / \rho^{1/2} \cdot g^{1/2} \cdot n^{1/2}$. If the half period is more than $R_d - R_c$, then any non zero solution $\varsigma(r)$ vanishes at most once on the interval $[R_c, R_d]$. In other words, if the following inequality hold

$$\pi \cdot \gamma^{1/2} \cdot \sin^{3/2} \alpha_g / \rho^{1/2} \cdot g^{1/2} \cdot n^{1/2} > R_d(1 - 1/n)$$ or
$$n - 1/n^{1/2} > \pi \cdot \gamma^{1/2} \cdot \sin^{3/2} \alpha_c / R_d \cdot \rho^{1/2} \cdot g^{1/2}$$

then any non zero solution $\varsigma(r)$ vanishes at most once on the interval $[R_c, R_d]$. Hence, according to [28] Chap. 11, the solution $\eta(r)$ of Jacobi equation which satisfies $\eta(R_d) = 0$ and $\eta'(R_d) = 1$ has only one zero on the interval $[R_c, R_d]$. This means that the Jacobi condition for weak minimum is satisfied [27].

ii). For obtaining the instability result remark now that for the coefficients of the Jacobi equation the following inequalities hold:
$$r \cdot \gamma / [1 + (z')^2]^{3/2} \leq R_d \cdot \gamma \cdot \cos^3 \alpha_c$$ and
$$\rho \cdot g \cdot r \geq \rho \cdot g \cdot \frac{R_d}{n}$$

Therefore the equation
$$\frac{d}{dr}(R_d \cdot \gamma \cdot \cos^3 \alpha_c \cdot \xi') + \rho \cdot g \cdot \xi \cdot R_d / n = 0$$

is a „Sturm type lower bound equation" [28] Chap.11 for the Jacobi equation. An arbitrary solution of the above „Sturm type lower bound equation" is given by
$$\xi(r) = A \cdot \sin(\omega \cdot r + \varphi)$$

Here $A$ and $\varphi$ are arbitrary real constants and $\omega^2 = \rho \cdot g / n \cdot \gamma \cdot \cos^3 \alpha_c$.

The period of any non zero solution $\xi(r)$ is $2\pi/\omega = 2\pi \cdot \gamma^{1/2} \cdot n^{1/2} \cos^{3/2} \alpha_c / \rho^{1/2} \cdot g^{1/2}$. If the period is less than $R_d - R_c$ then any non zero solution $\xi(r)$ vanishes at least twice on the interval $[R_c, R_d]$. In other words if the following inequalities hold :

$$2\pi \cdot \gamma^{1/2} \cdot n^{1/2} \cos^{3/2} \alpha_c / \rho^{1/2} \cdot g^{1/2} < R_d \cdot (1 - 1/n)$$ or
$$n - 1/n^{3/2} > 2\pi \cdot \gamma^{1/2} \cdot \cos^{3/2} \alpha_c / R_d \cdot \rho^{1/2} \cdot g^{1/2},$$

then any non zero solution $\xi(r)$ vanishes at least twice on the interval $[R_c, R_d]$. Hence according to [28] Chapter 11 the solution $\eta(r)$ of Jacobi equation which satisfies $\eta(R_d) = 0$ and $\eta(R_d') = 1$ has at least two zero on the interval $[R_c, R_d]$. This means that the Jacobi condition for weak minimum is not satisfied [27].

## ACKNOWLEDGEMENT

This research did not receive any specific grant from funding agencies in the public, commercial, or not-for-profit sectors.